\documentclass{llncs}
\usepackage{llncsdoc}
\usepackage{amsmath, amssymb}
\usepackage{amscd,amsfonts,amssymb, color}
\usepackage{latexsym, graphicx, pstcol,pst-grad, pst-text, pst-node, pst-tree}

\newtheorem{defin}{Definition}

\newtheorem{eg}{Example}
\newcommand{\mr}{\mathrm}
\newcommand{\sa}{self~assembly~}
\newcommand{\sg}{\Sigma}
\newcommand{\reg}{$REG$}
\newcommand{\ve}{\varepsilon}
\begin{document}
\title{A proposal to a generalised splicing with a self assembly approach}
\author{L Jeganathan \and R Rama \and Ritabrata Sengupta}
\institute{Department of Mathematics\\ Indian Institute of Technology\\Chennai 600 036, India\\ {\tt lj, ramar, rits@iitm.ac.in}}
\maketitle
\begin{abstract}
Theory of splicing is an abstract model of the recombinant behaviour of DNAs. In a splicing system, two strings to be spliced are taken from the same set and the splicing rule is from another set. Here we propose a generalised splicing (GS) model with three components, two strings from two languages and a splicing rule from third component. We propose a generalised self assembly (GSA) of strings. Two strings $u_1xv_1$ and $u_2xv_2$ self assemble over $x$ and generate $u_1xv_2$ and $u_2xv_1$. We study the relationship between GS and GSA. We study some classes of generalised splicing languages with the help of generalised self assembly.
\end{abstract}
\section{Introduction}
 Tom Head proposed \cite{head} an operation called `splicing', for describing
 the recombination of DNA sequences under the  application of restriction
 enzymes and ligases. Given two strings $u\alpha\beta v$ and $u'\alpha'\beta'
 v'$ over some alphabet  $V$ and a splicing rule
 $\alpha\#\beta\$\alpha'\#\beta'$, two strings $u\alpha\beta' v'$
 and $u'\alpha'\beta v$ are produced. The splicing rule
 $\alpha\#\beta\$\alpha'\#\beta'$ means that the first string is cut
 between $\alpha$ and $\beta$ and the second string is cut between
 $\alpha'$ and $\beta'$, and the fragments recombine crosswise.
 \par The splicing scheme (also written as
 H-scheme) is a pair $\sigma=(V,R)$ where $V$ is an alphabet and
 $R\subseteq V^*\#V^*\$V^*\#V^*$ is the set of splicing rules.
 Starting from a language, we generate a new language by the iterated
 application of splicing rules in $R$. Here $R$ can be infinite.
 Thus $R$ can be considered as a language over $V\cup\{\#,\$\}$.
 Splicing language (language generated by splicing) depends upon the
 class of the language (in the Chomskian hierarchy) to be
 spliced and the type of the splicing rules to be applied. The class
 of splicing language $H(FL_1,FL_2)$ is the set of strings generated
 by  taking any two strings from $FL_1$ and splicing them by the
 strings of $FL_2$. $FL_1$ and $FL_2$ can be any class of languages
 in the Chomskian hierarchy.
 Detailed investigations on computational power of splicing is found in \cite{dna}.
\par Theory of splicing is an abstract model of the recombinant behaviour of the DNAs. In a splicing system, the two strings to be spliced are taken from the same set and the splicing rule is from  another set. The reason for taking two strings from the same set  is, in the DNA recombination, both the objects to be spliced are DNAs.  For example, the splicing language in the class $H(FIN,REG)$ is the language generated by taking two strings from a finite language and using strings from a regular language as the splicing rules. Any general `cut' and `connection' model should include the cutting of two strings taken from two different languages. The strings spliced and the splicing rules have an effect on the language generated by the splicing process. In short, we view a splicing model as having three components, two strings from two languages as the first two components, and a splicing rule as the third component. Our proposal of a generalised splicing model (a formal definition of GS: Generalised splicing, is given in section 2 definition \ref{main})will be:
\[GS(L_1,L_2,L_3):=\{z_1,z_2: (x,y)\models_r (z_1,z_2),~x\in L_1, y\in L_2, r\in L_3\}.\]
  Instead of taking two strings from same language, as being done in the
 theory of splicing,  we take  them from two different languages. We cut them  by using  rules from a third language.  This means,
taking an arbitrary  word  $w_1(\in L_1)$ and an arbitrary  word from $w_2(\in L_2)$, we cut them by using an arbitrary rule of $L_3$. If $L_1=L_2$ in the generalised splicing model, we get the usual $H$-system.
\par
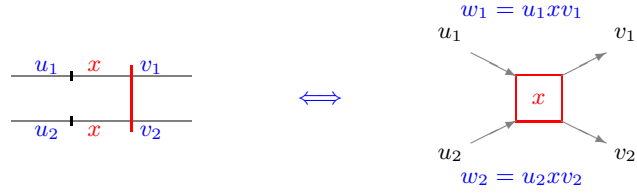
\begin{figure}[h]
\begin{center}
\parbox{5cm}{
\setlength{\unitlength}{2mm}
\begin{center}
\begin{picture}(20,10)(-10,-5)
{\gray\put(-6,1.5){\line(1,0){12}}
\put(-6,-1.5){\line(1,0){12}}}
\put(-2,1.8){\line(0,-1){0.6}}
\put(2,1.8){\line(0,-1){0.6}}
\put(-2,-1.2){\line(0,-1){0.6}}
\put(2,-1.2){\line(0,-1){0.6}}
{\red
\put(2,2.2){\line(0,-1){4.4}}}
{\blue\put(-5,1.8){$u_1$}\put(2,1.8){$v_1$}
\put(-5,-2.5){$u_2$}\put(2,-2.5){$v_2$}}
{\red \put(-2,1.8){$x$}
 \put(-2,-2.5){$x$}}
\end{picture}
\end{center}}
{\blue $\Longleftrightarrow$}
\parbox{5cm}{
\setlength{\unitlength}{2mm}
\begin{center}
\begin{picture}(20,10)(-10,-5)
{\red\put(-1.5,-1.5){\framebox(3,3){$x$}}}
\put(-4.5,-3){\gray{\vector(2,1){3}}}\put(-6.7,-4){$u_2$}
\put(-4.5,3){\gray{\vector(2,-1){3}}}\put(-6.7,4){$u_1$}
\put(1.5,1.5){\gray{\vector(2,1){3}}}\put(5,4){$v_1$}
\put(1.5,-1.5){\gray{\vector(2,-1){3}}}\put(5,-4){$v_2$}
\put(-5.2,5.5){{\blue $w_1=u_1xv_1$}}
 \put(-5.2,-5.6){{\blue $w_2=u_2xv_2$}}
\end{picture}
\end{center}}
\end{center}
\caption{Equivalence of generalised  splicing and \sa}
\label{assemble}
\end{figure}

 The motivation of the above proposal of a generalised theory of splicing comes from the self assembly of strings \cite{CPV}. Two strings $uv$ and $vw$ self assemble over $v$ and generate $uvw$. Here, the overlapping strings appear at the end of one string and at the beginning of the other. Then comes the question: What will be the generalisation if we do not restrict the overlapping strings to be in  the end (or the beginning) of the strings that participate in the assembling process. As an answer to the above question, we propose a generalised \sa (GSA) of two strings (definition \ref{main2}).
Two strings $u_1xv_1$ and $u_2xv_2$ self assemble over the sub-string $x$ and generate the strings  $u_1xv_2$ and $u_2xv_1$, as illustrated in the right hand side of the figure \ref{assemble}.  The generated words indicate that the $x$-\sa of $w_1$ and $w_2$ (\sa with $x$ as the overlapping string) is just a generalised  splicing of $w_1$ and $w_2$ with a splicing rule $x\#\$x\#$.
\par We take advantage of this equivalence of $GS$ and $GSA$ and plan to investigate the generalised splicing for some classes of languages in Chomskian hierarchy.
Since an investigation of the classes of languages under the generalised splicing model is going to be a  more complicated one, compared to the existing $H$-system in all sense, we narrow down the investigation of  the generalised splicing model by taking $L_3=V^+\cup\{(w_1,w_2):w_1\in L_1, w_2\in L_2\}$ ($V$ is the set of common symbols that appear in $L_1$ and $L_2$), which constitutes the set of splicing rules: a word $w\in V^+$ indicates that the splicing rule will be $w\#\$w\#$, where $\#$ and $\$$ have the usual meanings as in $H$-system; a pair of words $(w_1,w_2)\in L_3$ indicate that the splicing rule will be of the form $w_1\#\$w_2\#$. The very purpose of including the pair $(w_1,w_2)$ in $L_3$ is to include the words that are being spliced, in the set of words generated by the GS. The necessity of including the parent words is discussed at the end of section 2.
\par Though the whole theory of splicing can be rewritten with the generalised splicing system, nevertheless, in this paper, we investigate $GS(L_1, L_2, L_3)$ for $L_1, L_2\in \{REG, LIN, CF\}$ and $L_3$ is as given in the previous paragraph. For an investigation, we define the GSA of automata, regular grammar, linear grammar, context free grammar (apart from the GSA of two languages). In this paper, section 2 discusses the  definitions of GS and GSA. The subsequent sections discuss the generalised self assembly of finite languages, regular languages, linear languages and the context free languages. 
\section{Definitions}

Throughout this paper, we follow the terminologies and the notations as in \cite{AS}, \cite{dna}.
\begin{defin}[Generalised splicing scheme]\label{main} Generalised splicing scheme is defined as a triplet $\sigma_G=(V_1,$\\$V_2,R)$, where $V_1$, $V_2$ are alphabets, and $R\subseteq V_1^*\#V_1^*\$V_2^*\#V_2^*$. Here $R$ can be infinite, and $R$ is considered as a set of strings, hence a language. For a given $\sigma_G$, and a languages $L_1\subseteq V_1^*$ and $L_2\subseteq V_2^*$, we define
\[\sigma_G(L_1,L_2)=\{z_1,z_2:(x,y)\models_r (z_1,z_2),~\mbox{for~} x\in L_1, y\in L_2, r\in R\}.\]
Given three families $FL_1,~FL_2,~FL_3$; we define
\[GS(FL_1,FL_2,FL_3)=\{\sigma_G(L_1,L_2):L_1\in FL_1,L_2\in FL_2, R\in FL_3\},\]
i.e. $GS(FL_1,FL_2,FL_3)$ is the set of strings generated by splicing a language of $FL_1$, and a language of $FL_2$, by using a set of splicing rules in $FL_3$.
\end{defin}
\begin{note}
Whenever we refer `generalised splicing', we mean generalised 2-splicing.
\end{note}
\begin{defin}[Generalised self assembly]\label{main2}
Let $w_1\in L_1$,$w_2\in L_2$ be any two words. The generalised $x$-self assembly operation $GSA_x(w_1,w_2)$ over $(\ve\neq)x\in\mr{sub}(w_1)\cap\mr{sub}(w_2)$ is defined as follows:
\[GSA_x(w_1,w_2)=\{u_1xv_1,u_2xv_2,u_1xv_2,u_2xv_1:w_1=u_1xv_1,w_2=u_2xv_2,\}.\]
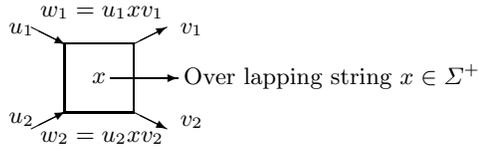
\begin{figure}[h]
\setlength{\unitlength}{1.5mm}
\begin{center}
\begin{picture}(20,10)(-10,-5)
\put(-3,-3){\framebox(6,6){$x$}}
\put(-6,-4.5){\vector(2,1){3}}\put(-8,-4){$u_2$}
\put(-6,4.5){\vector(2,-1){3}}\put(-8,4){$u_1$}
\put(3,3){\vector(2,1){3}}\put(7.2,4){$v_1$}
\put(3,-3){\vector(2,-1){3}}\put(7.2,-4.2){$v_2$}
\put(-5.2,5.5){$w_1=u_1xv_1$}
 \put(-5.2,-5.6){$w_2=u_2xv_2$}
 \put(1,0){\vector(1,0){6}}\put(7.5,-0.5){Over lapping string
 $x\in\Sigma^+$}
\end{picture}
\end{center}
\label{saw}
\caption{Super impose over the common sub-string $x$}
\end{figure}
\par The self assembled words are the words that are generated when we  trace from a left corner to a right corner in the  figure \ref{saw}.
Given any two languages $L_1$ and $L_2$, over the alphabet set $V_1$ and $V_2$ respectively, we define-
\[ GSA(w_1,w_2):=\bigcup_x GSA_x(w_1,w_2),\]
and
\[GSA(L_1,L_2):=\bigcup_{\substack{w_1\in L_1\\w_2\in L_2}} GSA(w_1,w_2).\]
\end{defin}
\par Though a \sa process will not include the parent words $w_1$ and $w_2$ (when $w_1\neq w_2$), in the above definition, we purposefully include the parent words for the sake of more clarity   of studying the GS through the GSA approach, i.e.  we plan to investigate $GS(L_1, L_2, L_3)$ where $L_3=V^+\cup\{(w_1,w_2): w_1\in L_1,  w_2\in L_2\}$ ($V$ is the set of common symbols that appear in $L_1$ and $L_2$). The pair $(w_1,w_2)$ in the set of splicing rules means that $w_1$ will be cut after $w_1$ and $w_2$ will be cut after $w_2$. Note that,the parent words $w_1$ and $w_2$ are included in $GS(w_1,w_2)$.

With the motivation given in section 1 and with the above two definitions, we have the following theorem.
\begin{theorem}\label[gsags]
Let $L_1$ and $L_2$ be any two languages. Let $V ~=~V_{L_1} \cap~~V_{L_2}$, where $V_{L_1}$ and $V_{L_2}$ are the alphabets of $V_{L_1}$ and $V_{L_1}$ respectively. Then
\[
GS(L_1, L_2,R) ~=~ GSA(L_1, L_2), 
\]
where
\[R ~=~ V^+ \cup \{(w_1, w_2):w \in L_1, w \in L_2\}\]
\end{theorem}

\section{Generalised Self assembly of finite languages}
This is the simplest and most trivial case. Suppose there are two finite languages $L_1$ and $L_2$, each containing $n_1$ and $n_2$ words respectively. Given any two words, there can be only finitely many common symbols between them. So only finitely many new words can be generated by self assembly. Since the parent languages are finite the end product $S(L_1,L_2)$ contains only finite number of words. Thus we get the following theorem:-
\begin{theorem}
Self assembly of two finite languages is finite. So we may write,
\[GSA(FIN,FIN)=FIN.\]
\end{theorem}
\section{Generalised Self assembly of regular languages}

In this section we shall investigate behaviour  of the \sa of two regular languages. We know that regular languages can be generated by regular grammar and are also accepted by a finite automata. We shall show that \sa of any two regular languages is regular. We shall prove it by both the automata and grammar approach.
\subsection{Generalised Self assembly of regular grammar}

In this section we shall describe: given any two regular grammars $G_1,G_2$ of languages $L_1$ and $L_2$ respectively, how to construct a grammar for the \sa~ language  $S(L_1,L_2)$.
\begin{defin}[Self assembly of REG grammars]\label{sag}
Let $G_i=(N_i,T_i,R_i,S_i), i=1,2$, be the regular grammars of languages $L_1=L(G_1)$ and $L_2=L(G_2)$,  where $N_i$'s are the set of non terminals, $N_1\cap N_2=\emptyset$, $T_i$'s are the set of terminals and $T_1\cap T_2\neq \emptyset$ (only then we can self assemble),  $S_i$'s are the  starting symbols and $R_i$'s are the rules respectively.\\

The generalised \sa of $G_1$ and $G_2$, written as $GSA(G_1,G_2)$ is defined as
\[G=(N_1\cup N_2\cup\{S\}, T_1\cup T_2, S,R),~S\notin N_1\cup N_2,\]
where $R$ includes the following rules:
\begin{enumerate}
\item $S\longrightarrow S_1,S\longrightarrow S_2$.
\item All the rules of $R_1$ and $R_2$.
\item For $a\in T_1\cap T_2$, for each pair of the rules $A\longrightarrow aB\in R_1$ and $A'\longrightarrow aB'\in R_2$, include the rules  $A\longrightarrow aB'$ and $A'\longrightarrow aB$ in $R$.
\end{enumerate}
\end{defin}
\begin{note}
\reg~ grammars are ones whose rules are of the form $A\longrightarrow aB$ or $A\longrightarrow a$, where $A$ is non-ter
minal and $a$ is a terminal. The two rules can be jointly expressed as $A\longrightarrow a\gamma$ where $\gamma$ is a non-terminal or $\gamma =\varepsilon$.
\end{note}

\begin{eg}
Let $G_1=(\{S_1\},\{a,b\},R_1=\{S_1\longrightarrow aS_1,~S_1\longrightarrow b\},S_1)$, and $G_2=(\{S_2\},\{a,b\},R_1=\{S_2\longrightarrow bS_2,~S_2\longrightarrow a\},S_1)$. $L_1=L(G_1)=a^*b$ and $L_2=L(G_2)=b^*a$. Then the GSA grammar is $G=(\{S,S_1,S_2\},\{a,b\},$\\$R,S)$, where the rules $R$ are given as
 \begin{align*}
 S\longrightarrow S_1 && S_1\longrightarrow aS_1|b|bS_2|a \\
 S\longrightarrow S_2 && S_2\longrightarrow aS_1|bS_2|a|b.
 \end{align*}
\end{eg}
\par Note that the language generated by $G$,  $L(G)$ will include the languages $L(G_1)$ and $L(G_2)$. Thus GSA of two regular grammars is again regular. In the same spirit of the above definition, we define GSA of linear grammars and GSA of context free grammars ( for this, we consider the Greibach normal form for CFG).
\begin{theorem}\label{sagt}
Let $G_1$ and $G_2$ be any two regular grammar. Then
\[L(GSA(G_1,G_2)=GSA(L(G_1),L(G_2)).\]
\end{theorem}
\proof
{\bf Part I:}\\
{\bf Case I} $w\in L(G_1)$ or $w\in L(G_2)$. It is trivial, since the rules $R_1$ and $R_2$ are included in $GSA(G_1,G_2)$.\\
{\bf Case II} $w\notin L(G_1)$ or $w\notin L(G_2)$. Let $w\in GSA(L(G_1),L(G_2))$. There exists $w_1\in L(G_1),~w_2\in L(G_2)$, $a\in \sg_{w_1}\cap \sg_{w_2}$, and $w=GSA(w_1,w_2)=uav$ such that $w_1=uau_1,~w_2=v_1av$, where $u\in\mr{prefix}(w_1),~v_1\in\mr{prefix}(w_2),~ u_1\in\mr{suffix}(w_1),~v\in\mr{suffix}(w_2)$. \\
Since $w_1\in L(G_1)$, there exists a sentential form
\[S_1\Rightarrow_{G_1}^*uA\Rightarrow uaB\Rightarrow_{G_1}^*uau_1:~~A\rightarrow aB\in R_1\]
for deriving $w_1=uau_1$. Similarly there exists a sentential form
\[S_2\Rightarrow_{G_2}^*v_1A'\Rightarrow v_1aB'\Rightarrow_{G_2}^*v_1av:~~A'\rightarrow aB'\in R_2\]
for deriving $w_2=v_1av$.
Since  $A\rightarrow aB\in R_1$ and         $A'\rightarrow aB'\in R_2$ implies that $A\rightarrow aB'\in R(GSA(G_1,G_2))$, we have the sentential form
\[S\Rightarrow_{GSA(G_1,G_2)}S_1\Rightarrow_{G_1}^*uA\Rightarrow_{GSA(G_1,G_2)} uaB'\Rightarrow_{G_2}^*uav'\]
i.e.
\[S\Rightarrow_{GSA(G_1,G_2)}uav=w.\]
Hence $w\in L(GSA(G_1,G_2)~\Rightarrow ~GSA(L(G_1),L(G_2))\subseteq L(GSA(G_1,G_2)$. \\
{\bf Part II:}\\

Let $w\in L(GSA(G_1,G_2))$. Without loss of generality, we assume that $w\notin L(G_1)$ and $L(G_2)$. \\

Since $w\in L(GSA(G_1,G_2))$, $w$ can be expressed as $w=uav$. So there exists a sentential form
\[S\Rightarrow_{GSA(G_1,G_2)}S_1\Rightarrow_{G_1}^* uA\Rightarrow_{GSA(G_1,G_2)} uaB'\Rightarrow_{G_2}^* uav\]
Since $A\rightarrow aB'\in R(GSA(G_1,G_2))$, but $\notin R_1,R_2$ (because $A,B\notin N_2$ and $A',B'\notin N_1$), there exists productions of the type $A\rightarrow aB\in R_1$ and   $A'\rightarrow aB'\in R_2$. \\

This implies
\[S_1\Rightarrow_{G_1}^*\Rightarrow uA\Rightarrow_{G_1}uaB\Rightarrow_{G_1}^*uax,~~~\mbox{using the production}~A\rightarrow aB\]
and
\[S_2\Rightarrow_{G_2}^*yA'\Rightarrow_{G_2} yaB'\Rightarrow_{G_2}^*yav,~~~\mbox{using the production}~A'\rightarrow aB'\]
This gives $\exists~uax\in L(G_1)$ and $yav\in L(G_2)$ corresponding to $w=uav\in GSA(L(G_1),L(G_2))$. \\

Hence $L(GSA(G_1,G_2)\subseteq GSA(L(G_1),L(G_2)).$ Hence the result.

\subsection{Generalised Self assembly of finite automata}
  If $L_1$ and $L_2$ any two \reg ~ languages, there exists two finite automatas $M_1$ and $M_2$ such that $L_1=L(M_1)$ and $L_2=L(M_2)$. While $L_1$ and $L_2$ can \sa~ by string overlapping, it is interesting to explore whether the corresponding automata self assemble to an automata $M$ such that the language of the self assembled automata is same as \sa~ of languages. If a word $w$ is accepted by a FA, every symbol $a$ in $w$ corresponds to an edge `$a$' in the transition diagram of the FA. This gives the idea that the FA's can be self assembled by the overlapping edge with the same level. Thus we have the following definition:
\begin{defin}[Generalised \sa of two FA's]\label{fa}
Let $M_1=(Q_1,V_1,\delta_1,q_1,F_1)$ and $M_2=(Q_2,V_2,\delta_2,$\\$
q_2,F_2)$ be two machines such that $V_1\cap V_2\neq\emptyset$. The generalised \sa of $M_1$ and $M_2$ written as $GSA(M_1,M_2)$ is defined as
\[M=(Q=Q_1\cup Q_2,V_1\cup V_2\cup\{\varepsilon\},\delta,q_0,F_1\cup F_2).\]
$\delta$ is defined as follows
\begin{enumerate}
\item $\delta(q_0,\varepsilon)=\{q_1,q_2\}$.
\item $\forall a\in V_1\cup V_2,q\in Q$
\[\delta(q,a)=\left\{\begin{array}{ll}
			\delta_1(q,a) & q\in Q_1\\
			\delta_2(q,a) & q\in Q_2
			\end{array}
	\right. \]
\item For every pair of transitions $\delta_1(q_i,a)=q_j$ and $\delta_1(q_i',a)=q_j',~q_1\in Q_1,~q_i'\in Q_2$, we include two new transition rules,
\begin{align*}
\delta(q_i,a)=q_j' && \delta(q_i',a)=q_j.
\end{align*}
\end{enumerate}
\end{defin}
\par Note that the language accepted by the GSA of $M_1$ and $M_2$ include $L(M_1)$ and $L(M_2)$.\\
\par It is observed that when $G_1$ and $G_2$ are regular grammars, we have
\[L(GSA(G_1,G_2))=L(GSA(M_1,M_2)),\]
where $L(G_1)=L(M_1)$ and $L(G_2)=L(M_2)$.
\par The idea behind the \sa~ of two FAs is the overlapping of the directed edge labelled with same symbol in the transition diagram of both the finite automatas. Every transition rules corresponds to a  directed edge in the  transition diagram. Let $\delta(q_i,a)=q_j$ and  $\delta(q_i',a)=q_j'$ be the transition in $M_1$ and $M_2$ respectively. In the \sa~ of $M_1$ and $M_2$, the directed edge in the transition diagram that corresponds to the above transition overlap:
\begin{figure}[h]
\begin{center}
\setlength{\unitlength}{3mm}
\begin{picture}(20,10)(-10,-5)
\multiput(-7,0)(14,0){2}{\circle{3}}
{\gray\qbezier(-6,1.1)(0,4.5)(6,1.1)\put(-0.2,2.8){\vector(1,0){1}}
\qbezier(-6,-1.1)(0,-4.5)(6,-1.1)\put(-0,-2.8){\vector(-1,0){1}}
\qbezier(-8,1.1)(-8.8,2)(-10,2)\put(-10,2){\vector(-1,0){0.5}}
\qbezier(8,1.1)(8.8,2)(10,2)\put(10,2){\vector(1,0){0.5}}
\qbezier(-8,-1.1)(-8.8,-2)(-10,-2)\put(-10,-2){\vector(-1,0){0.5}}
\qbezier(8,-1.1)(8.8,-2)(10,-2)\put(10,-2){\vector(1,0){0.5}}}
\put(-5.5,0){\vector(1,0){11}}
{\blue\put(0,.2){$a$}}
\put(-7.25,-0.2){$q_i$}
\put(6.75,-0.2){$q_j$}
\put(-2.8,-5){Machine $M_1$}
\end{picture}
\hspace*{2cm}
\setlength{\unitlength}{3mm}
\begin{picture}(20,10)(-10,-5)
\multiput(-7,0)(14,0){2}{\circle{3}}
{\gray\qbezier(-6,1.1)(0,4.5)(6,1.1)\put(-0.2,2.8){\vector(1,0){1}}
\qbezier(-6,-1.1)(0,-4.5)(6,-1.1)\put(-0,-2.8){\vector(-1,0){1}}
\qbezier(-8,1.1)(-8.8,2)(-10,2)\put(-10,2){\vector(-1,0){0.5}}
\qbezier(8,1.1)(8.8,2)(10,2)\put(10,2){\vector(1,0){0.5}}
\qbezier(-8,-1.1)(-8.8,-2)(-10,-2)\put(-10,-2){\vector(-1,0){0.5}}
\qbezier(8,-1.1)(8.8,-2)(10,-2)\put(10,-2){\vector(1,0){0.5}}}
\put(-5.5,0){\vector(1,0){11}}
{\blue\put(0,.2){$a$}}
\put(-7.25,-0.2){$q_i'$}
\put(6.75,-0.2){$q_j'$}
\put(-2.8,-5){Machine $M_2$}
\end{picture}\\
\setlength{\unitlength}{3mm}
\begin{picture}(20,10)(-10,-5)
\multiput(-7,0)(14,0){2}{\circle{3}}
{\gray\qbezier(-6,1.1)(0,4.5)(6,1.1)\put(-0.2,2.8){\vector(1,0){1}}
\qbezier(-6,-1.1)(0,-4.5)(6,-1.1)\put(-0,-2.8){\vector(-1,0){1}}
\qbezier(-8,1.1)(-8.8,2)(-10,2)\put(-10,2){\vector(-1,0){0.5}}
\qbezier(8,1.1)(8.8,2)(10,2)\put(10,2){\vector(1,0){0.5}}
\qbezier(-8,-1.1)(-8.8,-2)(-10,-2)\put(-10,-2){\vector(-1,0){0.5}}
\qbezier(8,-1.1)(8.8,-2)(10,-2)\put(10,-2){\vector(1,0){0.5}}}
\put(-5.5,0){\vector(1,0){11}}
{\blue\put(0,.2){$a$}}
\put(-8.2,-0.2){{\red$q_i$}}\put(-6.8,-0.2){{\red$q_i'$}}
\put(5.8,-0.2){{\red$q_j$}}\put(7.2,-0.2){{\red$q_j'$}}
\multiput(-7,-1.5)(14,0){2}{\line(0,1){3}}
\put(-2.8,-5){Machine $M$}
\end{picture}

\end{center}
\caption{Part of self assembled finite automata. Machine $M_1$ and $M_2$ are self assembled at the transitions edge $a$. The new FA $M$ is drawn to specifically highlight the assembled states.}
\label{ato}
\end{figure}
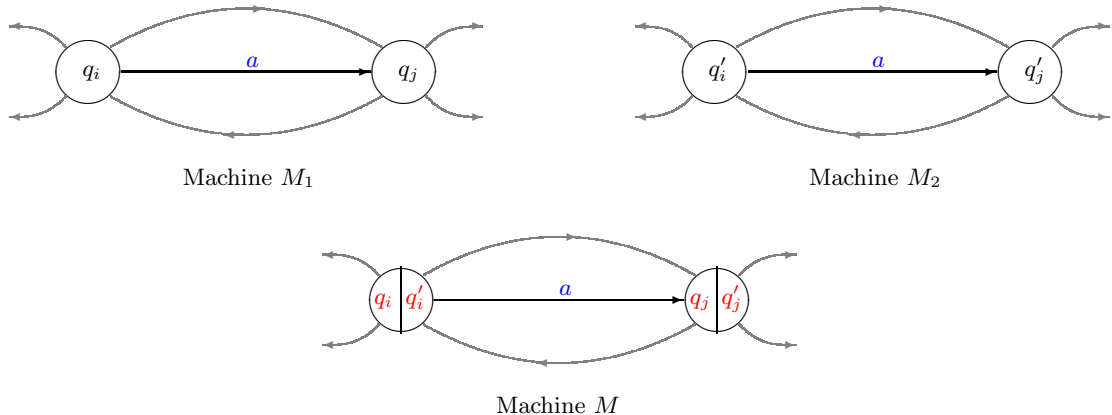
When the edges overlap, the states $q_i$ and $q_i'$ overlap.
To add more clarity, the figure \ref{ato} is drawn in a way that all the transitions are preserved.
\begin{theorem}\label{auto}
Let $M_1$ and $M_2$ be any two finite automatas. Then
\[L(GSA(M_1,M_2))=GSA(L(M_1),L(M_2)).\]
\end{theorem}
\proof
Without loss of generality, we assume that there is only one directed edge labeled $a$ in the transition diagram of $M_1$ and $M_2$, which can overlap. Further we can assume that all states of $M_1$ and $M_2$ are differently labeled.
\begin{description}
\item[Part I] Let $w\in L(GSA(M_1,M_2))$.
\begin{description}
\item[Case I] $w\in L(M_1))$ or $w\in L(M_2))$. Since $L(M_1),L(M_2)\subset GSA(L(M_1),L(M_2))$, we have $w\in GSA(L(M_1),L(M_2))$.
\item[Case II] $w\notin L(M_1))$ and $w\notin L(M_2))$ . There exists a path from $q_0$ to any one of the final states involving the edge $a$ the transition graph of $M$ such that the path preceding the edge $a$ is in $M_1$ (or in $M_2$), and the path succeeding the edge $a$ is in $M_2$ (or in $M_1$).\\
$\Rightarrow w=w_1aw_2$, where $w_1\mr{prefix}(x),~ x\in L(M_1)$ (or $w_1\mr{prefix}(x),~ x\in L(M_2)$) and $w_2\mr{suffix}(x),~ x\in L(M_2)$ (or $w_2\mr{prefix}(x),~ x\in L(M_1)$); i.e. $w_1$ is the labels of the path in $M_1$ (or in $M_2$), and  $w_2$ is the labels of the path in $M_2$ (or in $M_1$).\\
$\Rightarrow w$ can be written as the \sa~ of the words $w_1aw_1'$ and $w_2'aw_2$, where $w_1aw_1'\in L(M_1)$ and $w_2'aw_2\in L(M_2)$. \\
$\Rightarrow w\in GSA(L(M_1),L(M_2))$.\\
Hence  $L(GSA(M_1,M_2))\subset GSA(L(M_1),L(M_2))$.
\end{description}
\item[Part II] Let $w\in GSA(L(M_1),L(M_2))$. \\
$\Rightarrow w=GSA(x,y): ~x\in L(M_1),~y\in L(M_2)$.\\
$\Rightarrow w=w_1aw_2'$ or $w_2aw_1'$ where $x=w_1aw_1',~y=w_2aw_2'$.\\
$\Rightarrow$ There exists a path with label $w$ from $q_0$ to any one of the final states in the transition graph of $M$, involving the edge $a$.\\
$\Rightarrow w\in L(GSA(M_1,M_2))$.
\end{description}
Hence the result.\\

Combining the results above we get the following theorem.
\begin{theorem}
Generalised self assembly of two regular languages is regular. So we may write,
\[GSA(REG,REG)=REG.\]
\end{theorem}

We may also go a step further. For any $L_1\in FIN$ we can generate an automata $M_1$, in this way: for each word, make an automata which accepts only that word. All together  this will make a finite automata, with a unique starting symbol, which may take the empty string $\varepsilon$ and links to each of the individual automatas. Now given a regular language $L_2\in REG$, we have an automata $M_2$ accepting it. We can \sa them by the method described in  theorem \ref{auto}.  The resultant is again a finite automata. Since $L_2\subset GSA(L_1,L_2)$, by our construction, this automata also accepts infinite number of words. We can summarise this as:-
\begin{theorem}
Self assembly of  regular and finite languages is regular. So we may write,
\[GSA(FIN,REG)=REG.\]
\end{theorem}
\section{Generalised Self assembly of linear languages}
Linear languages (written as LIN) are the ones which are characterised by the following grammar rules.
\begin{align}
 X\longrightarrow P_1YP_2 && X\longrightarrow P,
\end{align}
where $X$ and $Y$ are non-terminals ($N$), and $P_1,~P_2, ~P_3$ are words over terminals($T$) \cite{AS}. If $P_1$ (resp. $P_2$) is   $\ve$ the grammar is called left-linear (resp. right-linear). Any linear language can be generated by right (or left) linear grammar. Also they are equivalent \cite{AS}. Hence for our purpose we convert all the grammars of the form of right-linear only, i.e. we are only considering rules of the form:
\begin{align*}
 X\longrightarrow P_1Y && X\longrightarrow P,
\end{align*}
where $P_1,P_2\in T^+$. Again we may further introduce new non-terminals, such that each rule is of either of the form:
\begin{align}\label{lin}
 X\longrightarrow aY && X\longrightarrow a,
\end{align}
where $a\in T\cup\{\ve\}$ and $Y\in N^+$.\\
{\bf Method for \sa of LIN grammar:}\\
 Now we use similar process as given in definition \ref{sag}. Suppose we have $L_1,L_2\in LIN$. We construct grammar     $G_i=(N_i,T_i,S_i,R_i), ~i=1,~2$ for them such that  $N_1\cap N_2=\emptyset$ and $R_i$'s are of the form of equation \ref{lin}. \\
Define a grammar $G=(N,T,S,R)$ where $N=N_1\cup N_2$, $T=T_1\cup T_2$, $S$ is the new starting symbol, and  the rules of $R$ are:
\begin{enumerate}
\item $S\longrightarrow S_1,S\longrightarrow S_2$.
\item All the rules of $R_1$ and $R_2$.
\item For $a\in T_1\cap T_2$, for each pair of the rules $A_1\longrightarrow a\gamma_1\in R_1$ and $A_2\longrightarrow a\gamma_2\in R_2$, include the rules  $A_1\longrightarrow a\gamma_2$ and $A_2\longrightarrow a\gamma_1$ in $R$, where $\gamma_1\in N_1^*$ and $\gamma_2\in N_2^*$.
\end{enumerate}
The analogous result of theorem \ref{sagt} follows the same line of argument. Thus we can also conclude that:
\begin{theorem}
Self assembly of two linear languages is linear; i.e.
\[GSA(LIN,LIN)=LIN.\]
\end{theorem}
\section{Generalised Self assembly of context free languages}
We self assemble CF grammars, and thus show that the \sa of two CF languages is again a CF language. Instead of using general grammar rules, we take the help of Greibach normal form \cite{HMU}. To use this, we can assume without loss of generality, that the parent languages are $\ve$ free. Now, in Greibach normal form each rule is of the form $A\longrightarrow a\gamma$, where $\gamma\in N^*$. We use exactly the same method used for linear grammar. Same lines of arguments give us:
\begin{theorem}
Generalised self assembly of two context free languages is context free; i.e.
\[GSA(CF,CF)=CF.\]
\end{theorem}
\section{Conclusion}
\par In all definitions of GSAs of languages, grammars (definition \ref{sag}) and FAs (definition \ref{fa}), the parent words are included in the words generated by the GSA. In fact, in any \sa process of   $w_1$ and $w_2$, $w_1$ $w_2$ will be generated only when $w_1=w_2$. But, in our definition of GSA, we prefer to include $w_1$ and $w_2$ (even if $w_1 \neq w_2$) in $GSA(w_1,w_2)$ with a purpose. Though we can define the GSA of grammars (as well as FAs) so that the parent words are not included in the words generated, the process will be highly complicated. The main purpose of this paper is just to study the generalised splicing in the \sa approach. For the sake of not loosing clarity of our approach in this study, we prefer to include the parent words in all our definitions, namely GS of languages, GSA of languages, and GSA of grammars.
\par
Thus, we have proved that $GS(FIN,FIN,R)=FIN$, $GS(REG,REG,R)=REG$, $GS(FIN,REG,R)=REG$, $GS(LIN,LIN,R)=LIN$ and $GS(CF,CF,R) =CF$, where $R$ is as mentioned as in Theorem 1.  This study can further be extended to study the other generalised splicing classes of languages.


\end{document}